\documentclass[11pt]{article}
\usepackage{graphicx}
\usepackage{epsfig}
\usepackage{subfigure}
\newcommand{\BABARPubYear}    {08}
\newcommand{\BABARConfNumber} {013}
\newcommand{\SLACPubNumber} {13353}

\input babarsym
\def\KKKz      {\ensuremath{K^+ K^- K^0}\xspace}
\def\KKKs      {\ensuremath{K^+ K^- \KS}\xspace}

\def\KKKspm      {\ensuremath{K^+ K^- {\KS}(\pip\pim)}\xspace}
\def\KKKszz      {\ensuremath{K^+ K^- {\KS}(\piz\piz)}\xspace}

\def\sPlot{\ensuremath{_s{\cal P}lot}\xspace}
\def\splot{\ensuremath{_s{\cal P}lot}\xspace}

\def\mKK   {\ensuremath{m_{\Kp\Km}}\xspace}

\def\betaeff {\ensuremath{\beta_{\mathit{eff}}}\xspace}
\def\Acp     {\ensuremath{{A}_{\CP}}\xspace}
\def\cosH    {\ensuremath{\cos \theta_H}\xspace}
\def\cosT    {\ensuremath{\cos \theta_{\rm T}}\xspace}

\def\fisher {\ensuremath{\mathcal{F}}\xspace}
\def\LowMass {Low-mass\xspace}
\def\HighMass {High-mass\xspace}
\def\hjphi {\ensuremath{\phi(1020)\xspace}}
\def\fzone {\ensuremath{f_0(980)\xspace}}
\def\hjX {\ensuremath{X(1550)\xspace}}

\long\def\inst#1{\par\nobreak\kern 4pt\nobreak
    {\it #1}\par\vskip 10pt plus 3pt minus 3pt}

\begin{document}
{\pagestyle{empty} 

\begin{flushright}
\babar-CONF-\BABARPubYear/\BABARConfNumber \\
SLAC-PUB-\SLACPubNumber \\
\end{flushright}

\par\vskip 5cm

\begin{center}
\Large \bf Measurement of {\boldmath$\CP$}-Violating Asymmetries in the {\boldmath $\Bz\to\Kp\Km\KS$} Dalitz Plot
\end{center}
\bigskip

\begin{center}
\large The \babar\ Collaboration\\
\mbox{ }\\
August 5, 2008
\end{center}
\bigskip \bigskip

\begin{center}
\large \bf Abstract
\end{center}
We present a preliminary measurement of \CP-violation parameters in the decay $\Bz \to \KKKs$,
using approximately 465 million $\BB$ events collected by the $\babar$ detector at SLAC. 
Reconstructing the neutral kaon as $\KS \to \pip\pim$ or  $\KS \to \piz\piz$,  we analyze the Dalitz plot 
distribution and measure fractions to intermediate states.  We extract \CP parameters from the asymmetries 
in amplitudes and phases between $\Bz$ and $\Bzb$ decays across the Dalitz plot. From a fit to the whole Dalitz 
plot, we measure $\betaeff= 0.44 \pm 0.07  \pm 0.02$, $\Acp= 0.03\pm 0.07 \pm 0.02$, where the first 
uncertainties are statistical and the second ones are systematic.  For decays to $\phi\KS$, we measure 
$\betaeff= 0.13 \pm 0.13 \pm 0.02$, $\Acp= 0.14 \pm 0.19  \pm 0.02$. For decays to $f_0\KS$, we measure 
$\betaeff= 0.15 \pm 0.13 \pm 0.03$, $\Acp= 0.01 \pm 0.26 \pm 0.07$.  From a fit to the region of the Dalitz 
plot with $\mKK>1.1\gevcc$, we measure $\betaeff= 0.52 \pm 0.08 \pm 0.03$, $\Acp= 0.05 \pm 0.09 \pm 0.04$.

\vfill
\begin{center}

Submitted to the 34$^{\rm th}$ International Conference on High-Energy Physics, ICHEP 08,\\
29 July---5 August 2008, Philadelphia, Pennsylvania.

\end{center}

\vspace{1.0cm}
\begin{center}
{\em Stanford Linear Accelerator Center, Stanford University, 
Stanford, CA 94309} \\ \vspace{0.1cm}\hrule\vspace{0.1cm}
Work supported in part by Department of Energy contract DE-AC02-76SF00515.
\end{center}

\newpage
}

\begin{center}
\small

The \babar\ Collaboration,
\bigskip

%
B.~Aubert,
M.~Bona,
Y.~Karyotakis,
J.~P.~Lees,
V.~Poireau,
E.~Prencipe,
X.~Prudent,
V.~Tisserand
\inst{Laboratoire de Physique des Particules, IN2P3/CNRS et Universit\'e de Savoie, F-74941 Annecy-Le-Vieux, France }
J.~Garra~Tico,
E.~Grauges
\inst{Universitat de Barcelona, Facultat de Fisica, Departament ECM, E-08028 Barcelona, Spain }
% ...

\end{center}\newpage

\section{INTRODUCTION}
\label{sec:Introduction}

We present a time-dependent analysis of the Dalitz plot (DP) in flavor tagged $\Bz\to
\KKKs$ decays, with the \KS reconstructed as $\KS \to \pip \pim$ or $\KS \to \piz \piz$
(unless otherwise stated, charge conjugates are implied throughout this paper).  
In the Standard Model (SM), these decays are dominated by 
$\b \to s\bar{s}s$ gluonic penguin amplitudes, with a single weak
phase.  Contributions from $b\to u \bar{q}q$ tree amplitudes,
proportional to the Cabibbo-Kobayashi-Maskawa (CKM) matrix element
$V_{ub}$ with a \CP-violating weak phase $\gamma$~\cite{ref:PDG}, are small, but may
depend on the position in the Dalitz plot.  In $\Bz\to\phi(\Kp\Km)\Kz$
decays the modification of the \CP asymmetry due to the presence of
suppressed tree amplitudes is at $\cal
O$(0.01)~\cite{Beneke:2005pu,Buchalla:2005us}, while at higher $\Kp\Km$ masses a larger
contribution at $\cal O$(0.1) is possible~\cite{Cheng:2005ug}.
Therefore, to very good precision, we also expect the direct \CP
asymmetry for these decays to be small in the SM.  The \CP asymmetry
in $\Bz \to \KKKs$ decay arises from the interference
of decays and $\Bz \leftrightarrow \Bzb$ mixing, with a relative phase
of $2\beta$.  The Unitarity Triangle angle $\beta$ has been measured
in  $\Bz\to [c\bar{c}]\Kz$ decays to be  
$\sin2\beta=0.685 \pm 0.032$~\cite{Aubert:2004zt,Abe:2005bt}.
Current direct measurements favor the solution of  $\beta=0.37$
over $\beta=1.20$ at the 98.3\% C.L.~\cite{Itoh:2005JspiKst,Krokovny:2006Dh0,Dalseno:2007DstDstKs,
Aubert:2005JPsiKpi,Aubert:2006BDsDsK,Aubert:2007BDh}.  Furthermore,
the $\beta=0.37$ solution is the only one consistent with all indirect constraints
~\cite{ref:CKMfit,ref:UTfit}.

The decay $\Bz \to \KKKs$ is one of the most promising processes with which to
search for physics beyond the SM.  Since  the leading  amplitudes
enter only at the one-loop level, additional contributions from heavy
non-SM particles may be of comparable size. If the amplitude from
heavy particles has a \CP-violating phase, the measured \CP-violation
parameters may differ from those expected in the SM.

Previous \babar\ measurements of the \CP asymmetry in $\Bz \to \KKKz$ decays
have been performed on $383\times 10^6$ \BB events~\cite{Previous}.  This 
analysis updates that previous result with a larger dataset.

\section{DATASET AND DETECTOR}
\label{sec:dataset}

The data used in this analysis were collected with the \babar\ detector
at the \pep2\ asymmetric-energy \B factory at SLAC. A total of 465
million \BB pairs were used.

The \babar\ detector is described in detail
elsewhere~\cite{ref:babar}.  Charged particle (track) momenta are
measured with a 5-layer double-sided silicon vertex tracker (SVT) and a
40-layer drift chamber (DCH) coaxial with a 1.5-T superconducting solenoidal
magnet.  Neutral cluster (photon) positions and energies are measured
with an electromagnetic calorimeter (EMC) consisting of 6580 CsI(Tl)
crystals.  Charged hadrons are identified with a detector of
internally reflected Cherenkov light (DIRC) and specific ionization
measurements (\dedx) in the tracking detectors (DCH, SVT).  Neutral hadrons that do not
interact in the EMC are identified with detectors, up to 15 layers
deep, in the flux return steel (IFR). 

In addition to the data collected by \babar, this analysis uses various samples of Monte Carlo (MC) events 
based on GEANT4~\cite{ref:geant4}. 
A sample of simulated \KKKs events using a full Dalitz plot model based on \babar's previous measurement is used to 
study signal events, while backgrounds from \B meson decays are studied using a separate sample of
simulated events.

\section{EVENT RECONSTRUCTION}
\label{sec:selection}

We reconstruct $\Bz \to \KKKs$ decays by combining two oppositely
charged tracks with a $\KS\to\pip\pim$ or $\KS\to\piz\piz$
candidate.  The \Kp and \Km tracks must have at least 12 measured DCH
coordinates, a minimum transverse momentum of
0.1~\gevc, and must originate from  the nominal beam spot.  Tracks
are identified as kaons using a likelihood ratio
that combines \dedx measured in the SVT and DCH with the Cherenkov
angle and number of photons measured in the DIRC.  The \Kpm candidates are required to be loosely compatible
with the kaon hypothesis if the \KpKm invariant mass is less than $1.1\gevcc$, while a tighter compatibility
is required in all other cases to further suppress background.

For all modes, the main source of background is random combinations of
particles produced in events of the type $e^+e^-\to q\bar{q}~(q=u,d,s,c)$ 
(continuum).  Additional background from decays of $B$
mesons to other final states (\BB background), with and without charm particles, is
estimated from MC events.

We use event-shape variables, computed in the center-of-mass (CM)
frame, to separate continuum events with a jet-like topology from the
more isotropic \B decays.  Continuum events are suppressed by
requiring the quantity $|\cosT|$ to be less than 0.9, 
 where $\theta_{\rm T}$ is the angle between the thrust axis calculated with the \B candidate's daughters and the thrust axis
formed from the other charged and neutral particles in the event.
Further discrimination comes from a Fisher discriminant (\fisher) based on 1) \cosT,
 2) 0th and 2nd order Legendre moments $\mathcal{L}_{i=0,2} = \sum_j p_j |cos(\theta_{j})|^i$,
where $j$ is all tracks and clusters not used to reconstruct the \B meson, $p_j$ is their momentum, and
$\theta_{j}$ is the angle to the \B thrust axis, and 3) the
magnitude of the cosine of the angle of the \B with respect to the collision axis $|\cos{\theta_{B}}|$.

In a small fraction of events, more than one \B candidate in a single event passes our selection
criteria.  In this case, a single best \B candidate is selected based on the \KS invariant mass and 
on the quality of the kaon tracks.

\B candidates are identified using two kinematic variables that separate
signal from continuum background. These are the beam-energy-substituted mass
$\mes \equiv \sqrt{ (s/2 + {\bf p}_{i}\cdot{\bf p}_{B})^{2}/E_{i}^{2}- {\bf
    p}^{2}_{B}}$, where $\sqrt{s}$ is the total \epem CM 
energy, $(E_{i},{\bf p}_{i})$ is the four-momentum of the initial
\epem system and ${\bf p}_{B}$ is the \B candidate momentum, both
measured in the laboratory frame, and $\Delta E \equiv E_{B} - \sqrt{s}/2$,
where $E_{B}$ is the \B candidate energy in the CM
frame.

\subsection{{\boldmath $\Bz \to \Kp\Km\KS$, $\KS\to\pip\pim$}}
\label{sec:kkkspm_selection}

\begin{figure}[ptb]
\center
\begin{tabular}{ll}
\includegraphics[height=5.5cm]{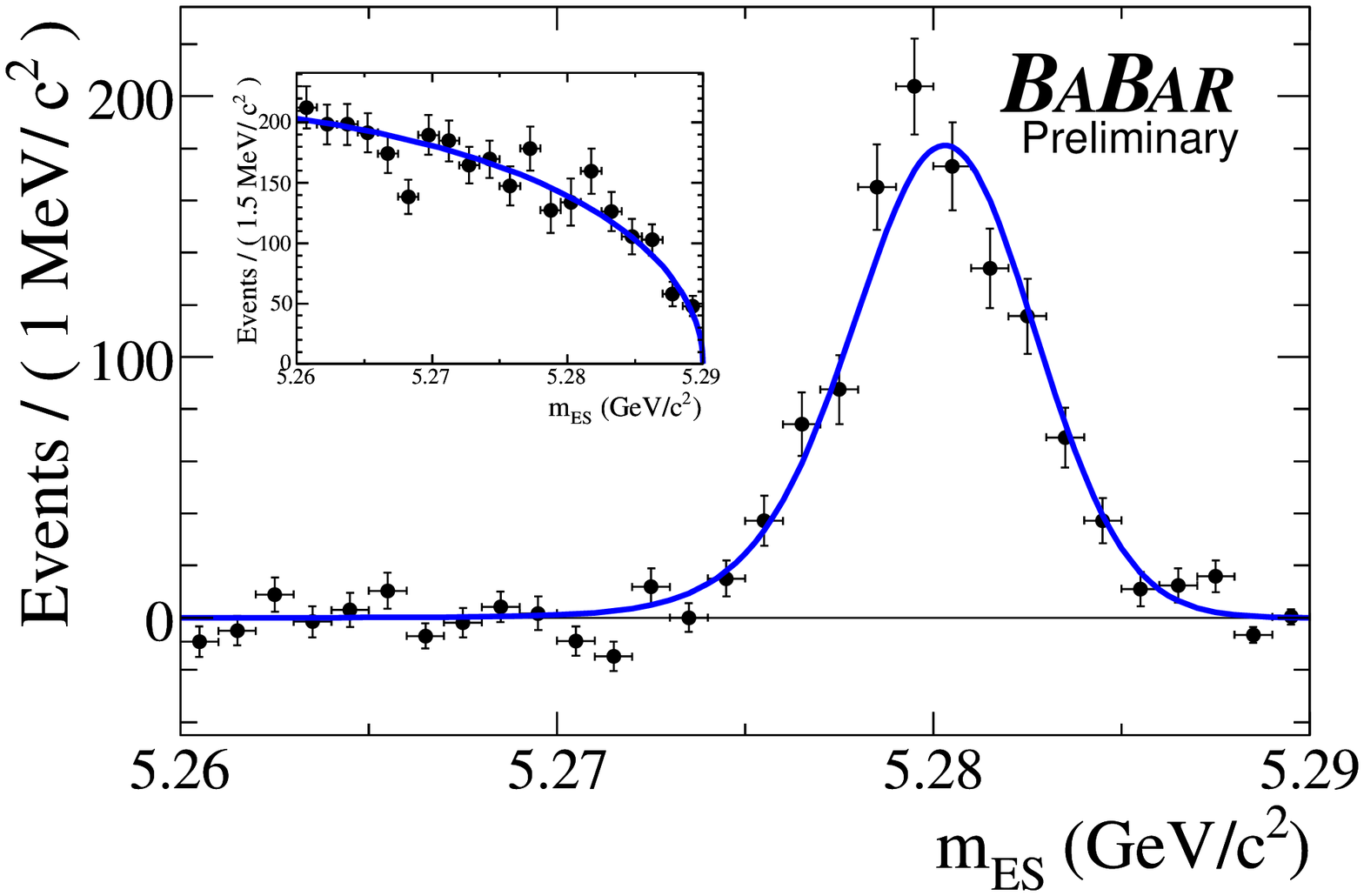} & \includegraphics[height=5.5cm]{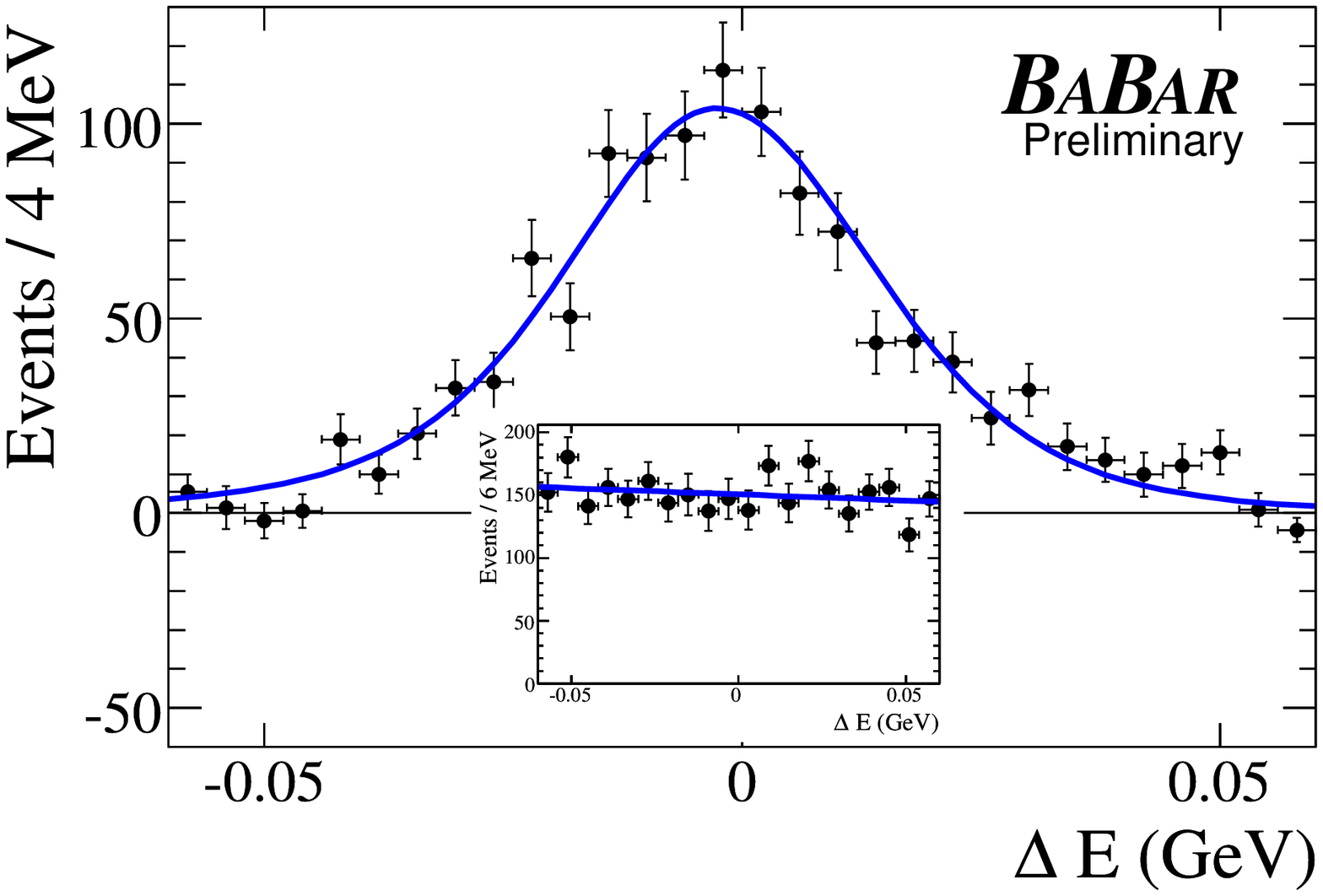} 
\end{tabular}
\caption{Distributions of kinematic variables (left) \mes\ and (right) \DeltaE for the \KKKspm sample.  The plots show
signal, with the continuum background shown in the insets. The points are data events weighted with the \sPlot\ 
technique, while the curves are the PDF shapes used in the ML fit (Sec. \ref{sec:Dalitz}).}
\label{fig:kkks+-_event_selection}
\end{figure}

For decays $\Bz \to \Kp\Km\KS$ and $\KS\to\pip\pim$, \KS candidates
are formed from oppositely charged tracks with an invariant mass within
$20~\mevcc$ of the \KS mass~\cite{ref:PDG}.  The \KS vertex is required to be
separated from the \Bz vertex by at
least $3\sigma$.  The angle $\alpha_{K_S}$ between the \KS momentum vector and the
vector connecting the \Bz and \KS vertices must satisfy $\cos\alpha_{K_S} >
0.999$. Distributions of the kinematic variables \mes and \DeltaE in data, 
for signal and background events calculated using the \splot event-weighting
method~\cite{ref:sPlots}, are shown in Fig.~\ref{fig:kkks+-_event_selection}.

\subsection{{\boldmath $\Bz \to \Kp\Km\KS$, $\KS\to\piz\piz$}}

For decays $\Bz \to \Kp\Km\KS$ and $\KS\to\piz\piz$, \KS candidates
are formed from two $\piz\to\gamma\gamma$ candidates.  Each of the
four photons must have $E_{\gamma} > 0.05 \gev$ and have a transverse
shower shape loosely consistent with an electromagnetic
shower. Additionally, we require each \piz candidate to satisfy $0.100
< m_{\gamma\gamma} < 0.155 \gevcc$. The resulting $\KS\to\piz\piz$
mass is required to satisfy $0.4776 < m_{\piz\piz} < 0.5276~\gevcc$. A \KS
mass constraint is then applied for the reconstruction of the \Bz
candidate.

\begin{figure}[ptb]
\center
\begin{tabular}{ll}
\includegraphics[height=5.5cm]{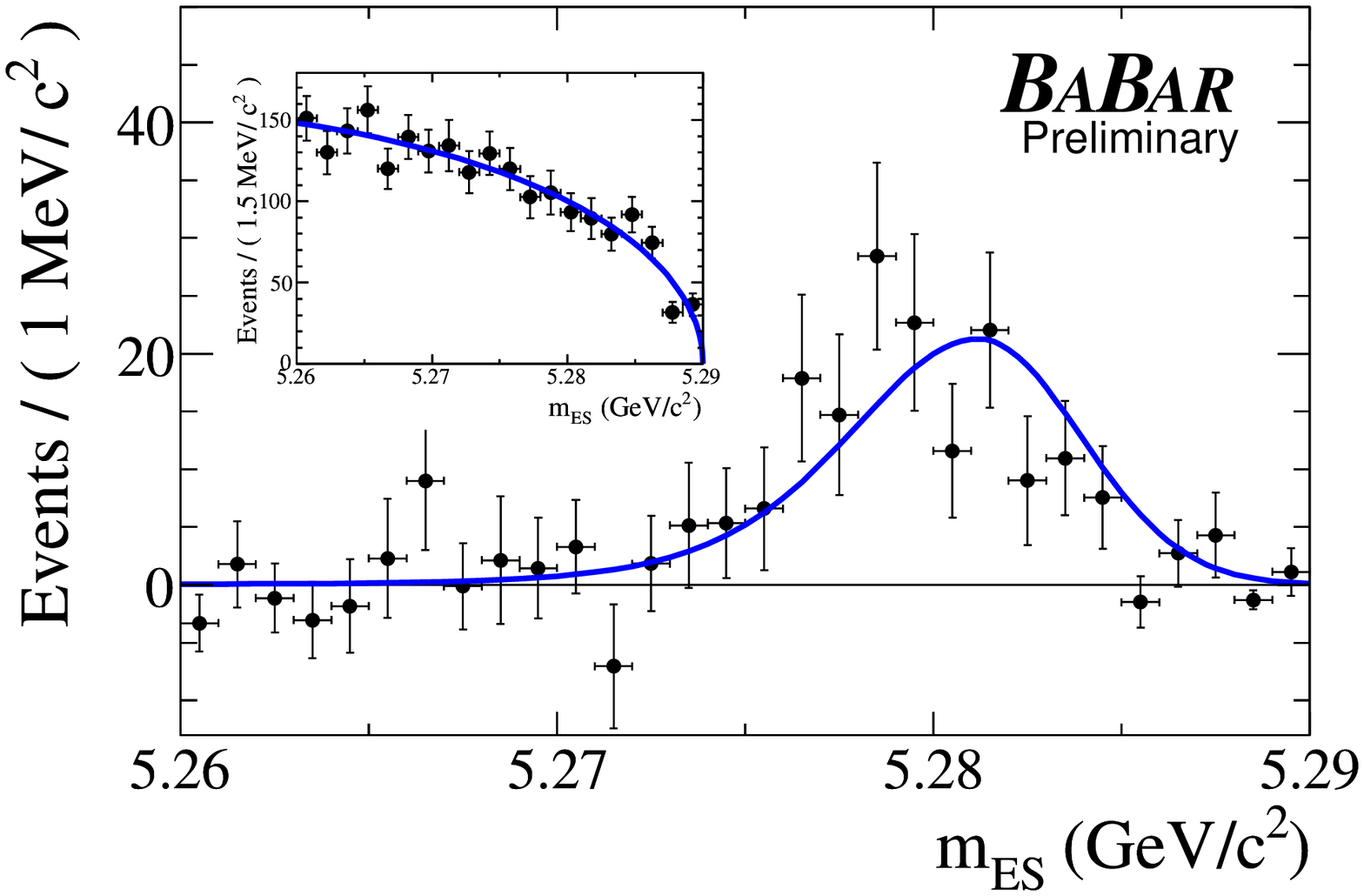} & \includegraphics[height=5.5cm]{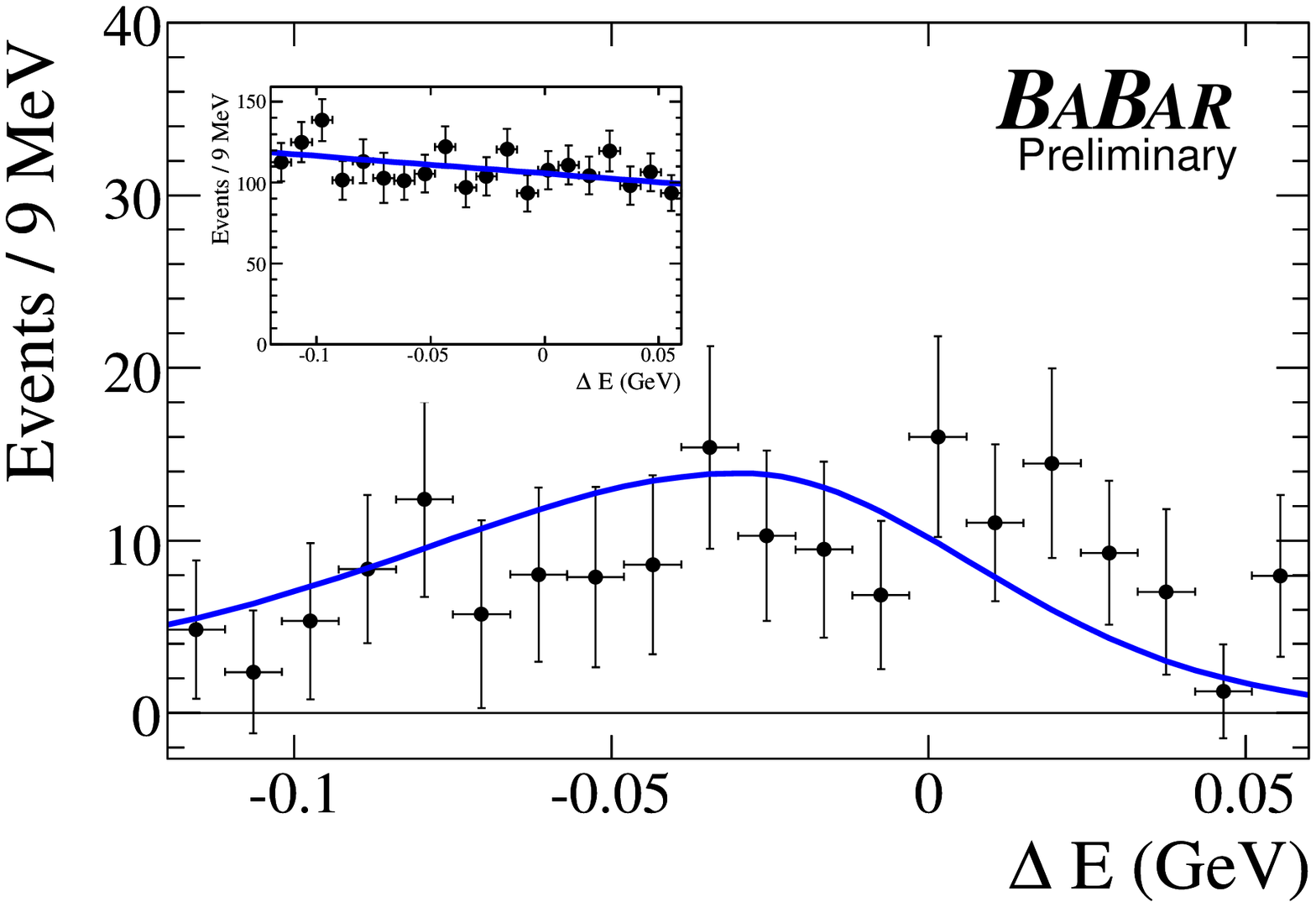} 
\end{tabular}
\caption{Distributions of kinematic variables (left) \mes\ and (right) \DeltaE for the \KKKszz sample.  The plots show
signal, with the continuum background shown in the insets. The points are data events weighted with the \sPlot\ 
technique, while the curves are the PDF shapes used in the ML fit (Sec. \ref{sec:Dalitz}).}
\label{fig:kkks00_event_selection}
\end{figure}

The kinematic variables \mes and \DeltaE are formed for each candidate
as in Sec.~\ref{sec:selection}. Distributions of these
variables in data, for signal and background events calculated using the \splot event-weighting
method, are shown in Fig.~\ref{fig:kkks00_event_selection}.
Note that the mean of the signal \DeltaE distribution is shifted from zero due to energy leakage in the EMC.

\section{ANALYSIS OF THE DALITZ PLOT}
\label{sec:Dalitz}
Four-momentum conservation in a three-body decay gives the relation
$ M^2_{\Bz} + m^2_{1} + m^2_{2} +
m^2_{3} ~=~ m^2_{12} + m^2_{13} + m^2_{23}$, where
$m^2_{ij}=(p_i+p_j)^2$ is the square of the invariant mass of a
daughter pair.  This constraint leaves a choice of two independent
Dalitz plot variables to describe the decay dynamics of a spin-zero
particle. In this analysis we choose the $\Kp\Km$ invariant mass \mKK
and the cosine of the helicity angle between the $\Kp$ and the $\KS$ in the 
$\Kp\Km$ center-of-mass frame,  \cosH. 

\begin{figure}[ptb]
\center
\epsfig{file=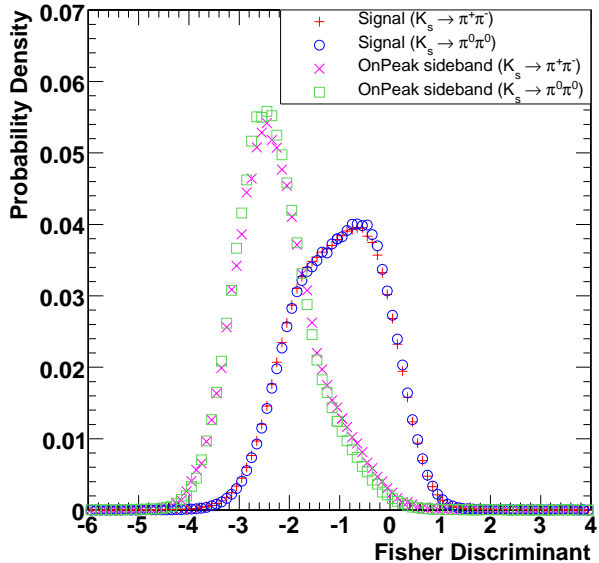, width=0.5\textwidth}\\
\caption{Fisher discriminant distributions for signal and continuum background and for \KKKspm sample and \KKKszz sample. 
  Distributions are normalized to unit area.}
\label{fig:Fisher}
\end{figure}

We perform an extended maximum likelihood fit to the measured time dependent Dalitz plot distribution. We first fit on the whole DP, then fit on the $\mKK>1.1\gevcc$ range (\HighMass), then fit on the $\mKK<1.1\gevcc$ range (\LowMass). All fits are performed on the combined \KKKspm and \KKKszz samples simultaneously.
The likelihood function ${\mathcal L}$ 
for each subsample is defined as
\begin{equation}
{\mathcal L} = \exp{\left(-\sum_{i}n_{i}\right)}
\prod_{j}\left[\sum_{i}n_{i}{\mathcal P}_{i,j}\right]
\label{eq::ml}
\end{equation}
where $i$ labels the different signal and background components, $j$ runs over all events in the sample, and $n_i$ is the event yield for events of the $i$-th component. The probability density function (PDF) ${\mathcal P}_i$ of each component is defined as 
\begin{equation}
\label{eq::mlprod}
{\mathcal P}_i \equiv {\mathcal P}_i(\mes) \cdot {\mathcal P}_i(\DeltaE) \cdot {\mathcal P}_i(\fisher) \cdot {\mathcal P}_{DP,i}(\mKK, \cosH, \deltat, q_{tag})\otimes {\cal R}_i(\deltat, \sigma_{\deltat}),
\end{equation}
where $q_{tag}$ is the flavor of the tagged \B (1 for \Bz and -1 for \Bzb), and $\deltat=t_{rec}-t_{tag}$ is the
difference of the proper decay times of the two
$B$-mesons in the \FourS decay. $\sigma_{\deltat}$ is the error on \deltat, and $\cal R$ is the \deltat\ resolution function determined from a high statistics independent sample~\cite{Aubert:2004zt}. 
For the purpose of calculating the DP coordinates
\mKK and \cosH, we refit the \B candidates applying a \B mass constraint.  This ensures that the \B candidates are reconstructed within the DP boundary.
The Fisher discriminant PDF, ${\mathcal P}(\fisher)$, is only used in the \LowMass fit (see Sec.~\ref{sec:Physics}).  Because
the Fisher discriminant is highly correlated with the position on the DP, we do not use the Fisher discriminant PDF for the fit to the whole DP or for the \HighMass fit.
 The Fisher distributions are shown in Fig.~\ref{fig:Fisher}. The PDFs for the individual fit components are described in more detail below.

\subsection{Background in the Time-Dependent Dalitz Plot}

We have two background components in our fit: continuum and \BB background. For the continuum background component,
we use the ARGUS function~\cite{Albrecht:1990cs} for ${\mathcal P}(\mes)$, and linear polynomial functions
for ${\mathcal P}(\DeltaE)$. The \deltat distribution is described by a double-Gaussian resolution function
convolved with a PDF of the following form:
\begin{equation}
	{\cal P}(\deltat) = f_{prompt} \delta(\deltat) + (1-f_{prompt}) e^{-|\deltat|/\tau_{bg}},
\end{equation}
which allows for background decays with both zero and non-zero lifetimes. The Dalitz plot for the continuum 
background is parameterized using a two-dimensional histogram PDF in the variables \mKK and \cosH. The histogram 
is filled with candidates from the region $5.2 < \mes < 5.26~\gevcc$.

We estimate the amount of $\BB$ background from Monte Carlo events. The \BB\ background is
almost purely combinatorial and is a few percent of the total background. 
In the \KKKspm mode,  the \mes and \DeltaE PDFs for the \BB backgrounds are
parameterized with the same functional forms as the continuum backgrounds. 
Due to non-negligible correlation between \mes and \DeltaE for \BB background in 
the \KKKszz mode, we construct a two-dimensional smoothed histogram PDF in those variables.
The \deltat\ distribution is described with a PDF similar to the continuum backgrounds, but we
also allow for the possibility that the non-zero lifetime component has a time-dependent \CP asymmetry 
proportional to $\sin\deltamd\deltat$ or $\cos\deltamd\deltat$, where \deltamd is the mixing frequency of
the \Bz meson. These asymmetries are set to zero in the nominal fit, but are varied as a systematic uncertainty.
The Dalitz plot is described using a two-dimensional histogram PDF in a manner similar to the continuum 
backgrounds.

\subsection{Signal Decays in the Time-Dependent Dalitz Plot}
\label{sec:tddp}

The signal components of the PDFs for ${\mathcal P}(\mes)$ and ${\mathcal P}(\DeltaE)$ are parameterized using 
modified Gaussian distributions:
${\mathcal P}(x) \propto \exp [ - (x - x_0)^2/(2 \sigma_{\pm}^2 + \alpha_{\pm} (x- x_0)^2) ].$
We determine the parameters $x_0$, $\sigma_+$, $\sigma_-$, $\alpha_+$, and $\alpha_-$ using MC events, and fix them 
in fits to data. For $x<x_0$ ($x>x_0$), the parameters $\sigma_-, \alpha_-$ ($\sigma_+, \alpha_+$) are used.

For signal events, the time-dependence is a function of location in the DP.
When the flavor of the tagged \B $q_{tag}$, and the difference of
the proper decay times \deltat, are measured,
the time- and flavor-dependent decay rate over the Dalitz plot can be
written as
\begin{eqnarray}
d\Gamma =\frac{1}{(2\pi)^3}\frac{1}{32 M_{\Bz}^3}  \frac{e^{-|\deltat|/\tau_{\Bz}}}{4\tau_{\Bz}} &\times& 
        \Big[~ \left | {\cal A} \right |^2 + \left | \bar{ {\cal A} } \right |^2 \nonumber 
        + q_{tag}~(1-2 w) ~2 Im \left ( e^{-2i\cdot\beta} \bar{\cal A} {\cal A}^*  \right ) \sin\deltamd\deltat \\ 
        && -~ q_{tag}~(1-2 w) ~\left (\left | {\cal A} \right |^2 - \left | \bar{ {\cal A} } \right |^2 \right ) \cos\deltamd\deltat 
        ~\Big ], 
\label{eq::dalitz_plot_rate}
\end{eqnarray}
where $q_{tag} = +1(-1)$ when the other \B meson is identified as a \Bz(\Bzb) using a neural network 
technique~\cite{Aubert:2004zt}. The parameter $w$ is the fraction of events in which the \B meson is mistagged with 
the incorrect flavor, and the parameter $\beta$ is the CKM
angle $\beta$, coming from \Bz-\Bzb mixing. 
Approximately 75\% of the signal events have tagging information
and contribute to the measurement of CP violation parameters. After
accounting for the mistag rate, the effective tagging efficiency 
is $(31.2\pm 0.3)\%$. 
Events without tagging information are assigned a mistag rate of $w=0.5$, and are included in the fit as they contribute
to the determination of the Dalitz plot parameters.
Decay amplitudes $\cal A$ and $\bar{\cal A}$ are defined in
(\ref{eq:A}) and (\ref{eq:Abar}) below. 
$M_{\Bz}$, $\tau_{\Bz}$, and \deltamd are
the mass, lifetime, and mixing frequency of the \Bz meson, respectively~\cite{ref:PDG}.

The PDF for the Dalitz plot rate takes the form
\begin{eqnarray}
        {\cal P}_{DP} \propto d\Gamma(\mKK,\cos\theta_H,\deltat, q_{tag})  \cdot \varepsilon(\mKK,\cos\theta_H)  \cdot |J| 
                        \otimes {\cal R}(\deltat, \sigma_{\deltat}),
\end{eqnarray}
where $|J(\mKK)| = (2 \mKK)(2 q p)$ is the Jacobian of the transformation $(m^2_{\Kp\Km},m^2_{\Kp\KS}) \leftrightarrow (m_{\Kp\Km},\cos\theta_H)$, and is
given in terms of the charged kaon momentum $q$ and neutral kaon momentum $p$, in the $\Kp\Km$ frame.  
The efficiency $\varepsilon$ is calculated from high-statistics samples of simulated events and depends on the position on the Dalitz plot.

The amplitude $\cal A$ ($\bar{\cal A}$) for the decay $\Bz\to\Kp\Km\KS$ ($\Bzb\to\Km\Kp\overline{\KS}$) is, in our isobar model, 
written as a sum of decays through intermediate resonances:
\begin{eqnarray}
        {\cal A} &=& \sum\limits_r c_r (1+b_r) e^{i (\phi_r + \delta_r)} \cdot f_r, \hspace{1cm}\mathrm{and} \label{eq:A} \\
        \bar{\cal A}&=& \sum\limits_r c_r (1-b_r) e^{i (\phi_r - \delta_r)} \cdot \bar{f}_r. \label{eq:Abar}
\end{eqnarray}
The isobar coefficients $c_r$ and $\phi_r$ are the magnitude and phase of the
amplitude of component $r$, and we allow for different isobar
coefficients for $\Bz$ and $\Bzb$ decays through the asymmetry
parameters $b_r$ and $\delta_r$.  The
function $f_r = F_r \times T_r \times Z_r$ describes the dynamic
properties of a resonance $r$, where $F_r$ is the form-factor for the
resonance decay vertex, $T_r$ is the resonant mass-lineshape, and
$Z_r$ describes the angular distribution in the
decay~\cite{blatt,Zemach:1963bc}.

Our model includes the $\phi(1020)$, for which we use the  Blatt-Weisskopf centrifugal barrier factor
$F_r=1/\sqrt{1+(Rq)^2}$~\cite{blatt},  where $q$ is the daughter momentum in the resonance frame,
and $R$ is the effective meson radius, taken to be $R=1.5~\gev^{-1}~(0.3~\fm)$. 
For the scalar decays included in our model ($f_0(980)$, $\hjX$, and $\chi_{c0}$), we use a constant form-factor.
Note that we have omitted a similar centrifugal factor for the $\Bz$ decay vertex into the $\phi\Kz$ intermediate state
since its effect is negligible due to the small width of the $\phi(1020)$ resonance.

The angular distribution is constant for scalar decays, whereas for vector decays $Z\sim \vec{q} \cdot \vec{p}$, 
where $\vec{q}$ is the momentum of the resonant daughter, and $\vec{p}$ is the momentum of the third particle in 
the resonance frame. 
We describe the line-shape for the $\phi(1020)$, $\hjX$, and $\chi_{c0}$  using the relativistic Breit-Wigner function 
\begin{equation}
        T(m) = \frac{1}{m^2_r - \mKK^2 - i m_r \Gamma(m)},
\end{equation}
where $m_r$ is the resonance pole mass. The mass-dependent width is given as
$
\Gamma(\mKK) =\Gamma_r  \left ( q/q_r\right )^{2L+1} \left ( m_r / \mKK  \right ) \left ( F_r(q)/F_r(q_r) \right )^2,
$
where $L$ is the resonance spin and $q=q_r$ when $\mKK=m_r$.
For the $\phi(1020)$ and $\chi_{c0}$ parameters, we use average measurements~\cite{ref:PDG}. 
The $\hjX$ is less well-established.
Previous Dalitz plot analyses of $\Bp\to\Kp\Kp\Km$~\cite{Garmash:2004wa,Aubert:2006nu} and
$\Bz\to\Kp\Km\Kz$ decays~\cite{Aubert:2005kd}  report observations of a scalar resonance at around 1.5~\gevcc.
The scalar nature has been confirmed by partial-wave analyses~\cite{Aubert:2005ja,Aubert:2006nu}.
However, previous measurements report inconsistent resonant widths: $0.145\pm 0.029$~\gevcc~\cite{Garmash:2004wa} and
$0.257 \pm 0.033$~\gevcc~\cite{Aubert:2006nu}. Branching fractions also disagree, so the nature of this component is still unclear~\cite{Minkowski:2004xf}.
In our nominal fit, we take the resonance parameters from Ref.~\cite{Aubert:2006nu}, which is based on
a larger sample of $\BB$ decays than Ref.~\cite{Garmash:2004wa}, and consider the narrower width given in the latter in the systematic error studies.

The $f_0(980)$ resonance is described with the coupled-channel (Flatt\'e) function
\begin{equation}
        T(\mKK)    =  \frac{1 }{   m^2_r - \mKK^2 - i m_r ( \rho_{K} g_{K} + \rho_{\pi} g_{\pi}  )     }, 
\end{equation}
where  $\rho_K (\mKK) =2 \sqrt{ 1 - 4 m^2_{K}/\mKK^2 }$, $\rho_\pi (\mKK) =2 \sqrt{ 1 - 4 m^2_{\pi}/\mKK^2 }$, and
the coupling strengths for the $KK$ and $\pi\pi$ channels are 
taken as $g_\pi=0.165\pm0.018$~\gevcc, $g_K/g_\pi=4.21 \pm 0.33$, 
and $m_r=0.965 \pm 0.010 $~\gevcc~\cite{Ablikim:2004wn}.

In addition to resonant decays, we include non-resonant amplitudes. 
Existing models consider contributions from contact terms or higher-resonance
tails~\cite{Cheng:2002qu,Fajfer:2004cx,Cheng:2005ug}, but they do not capture features observed in data.
We rely on a phenomenological parameterization~\cite{Garmash:2004wa} and 
describe the non-resonant terms as
\begin{equation}
         {\cal A}_{NR} (\bar{\cal A}_{NR}) =  \left ( c_{12} e^{i\phi_{12}} e^{-\alpha m^2_{12}}  
             +   c_{13} e^{i\phi_{13}} e^{-\alpha m^2_{13}}  
             +    c_{23} e^{i\phi_{23}} e^{-\alpha m^2_{23}} \right )  
	     \cdot (1 \pm b_{NR}) \cdot  e^{\pm i\delta_{NR}}, \label{eq:nr}
\end{equation}
where 1,2,3 denote the three daughter particles of the \B meson. 
The slope of the exponential function is consistent among previous measurements in both neutral and
charged $B$ decays into three kaons~\cite{Garmash:2004wa,Aubert:2006nu,Aubert:2005kd}, and we use $\alpha = 0.14 \pm 0.02~\gev^{-2} \cdot c^4$.

We compute the direct \CP-asymmetry parameters for resonance $r$ from the asymmetries in amplitudes ($b_r$) 
and phases ($\delta_r$) given in Eqs.~(\ref{eq:A}, \ref{eq:Abar}). We define the rate asymmetry as
\begin{equation}
        \Acp(r)=\frac{|\bar{\cal A}_r|^2 - |{\cal A}_r|^2}{|\bar{\cal A}_r|^2 + |{\cal A}_r|^2} =\frac{-2b_r}{1+b_r^2},
\label{eq:Acp}
\end{equation}
and $\betaeff (r) = \beta + \delta_r$ is defined as the total phase asymmetry.  These asymmetries are related to the
\CP asymmetry parameters $C$ and $-\eta S$ using the approximations
\begin{eqnarray}
        C_r \approx - \Acp(r), \hspace{1cm}\mathrm{and} \label{eq:C} \\
        -\eta_r S_r \approx \frac{1-b_r^2}{1+b_r^2} \sin(2 \betaeff (r) ), \label{eq:etaS}
\end{eqnarray}
where $\eta_r$ is the \CP eigenvalue of the final state.  The fraction for resonance $r$ is computed as
\begin{equation}
	{\cal F}_r ~=~ \frac{ \int d\cos\theta_H ~d\mKK \cdot |J| \cdot (|{\cal A}_r |^2+|\bar{\cal A}_r|^2)    }
		     { \int  d\cos\theta_H ~d\mKK \cdot |J| \cdot (|{\cal A} |^2+|\bar{\cal A}|^2)    }.
\end{equation}
The sum of the fractions can be different from unity due to interference between the isobars.

In addition to the previously mentioned resonances, the decays $\Bz \to \Dp\Km~(\Dp \to \Kp\KS)$ and 
$\Bz \to \Ds\Km~(\Ds \to \Kp\KS)$ are also counted as signal. We include non-interfering amplitudes for 
these modes in our Dalitz plot
model, parameterizing the $D_{(s)}$ mesons on the Dalitz plot as Gaussian
distributions with widths taken from studies of simulated events. The parameters $b_r$ and $\delta_r$ are 
fixed to zero for the decays  $\Bz \to \Dp\Km$, $\Bz \to \Ds\Km$, and $\Bz \to \chi_{c0}\KS$ throughout
this analysis.

\section{RESULTS}
\label{sec:Physics}

In order to determine parameters of the Dalitz plot model,
we perform three fits: 1) whole DP fit, 2) \LowMass ($\mKK<1.1\gevcc$) region fit, and 3) \HighMass ($\mKK>1.1\gevcc$) region fit.

\subsection{The whole Dalitz Plot fit}
\begin{figure}[ptb]
\center
\begin{tabular}{c}
\includegraphics[height=6.5cm]{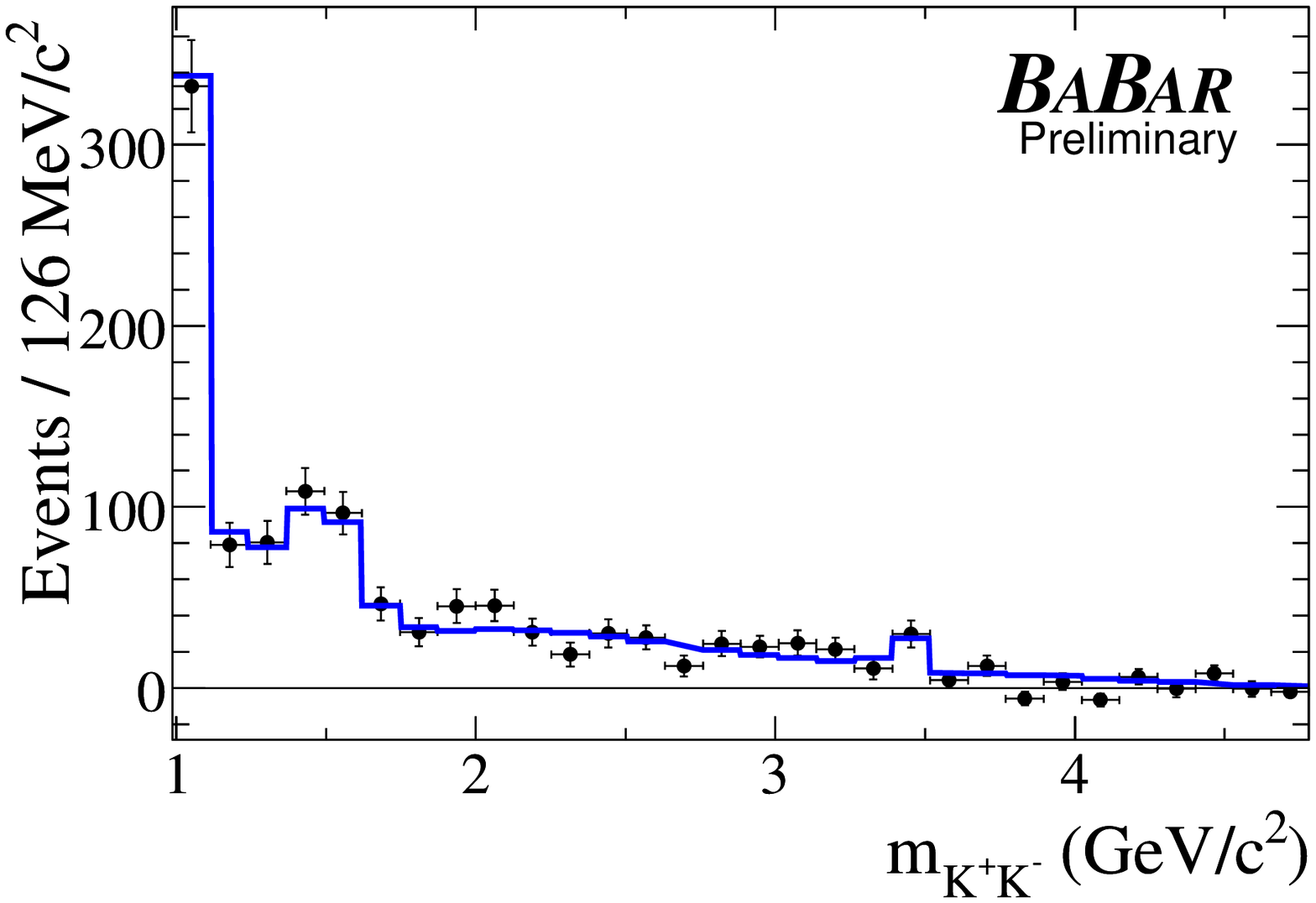} 
\vspace{-.5cm}
\end{tabular}
\caption{For the whole DP region fit, the distribution of the Dalitz plot variable \mKK for 
signal-weighted data events (points) compared with the fit PDF in the \KKKspm mode. }
\label{fig:DV-wholeDP}
\end{figure}

\begin{figure}[ptb]
\center
\begin{tabular}{ll}
\includegraphics[width=8cm,height=4cm]{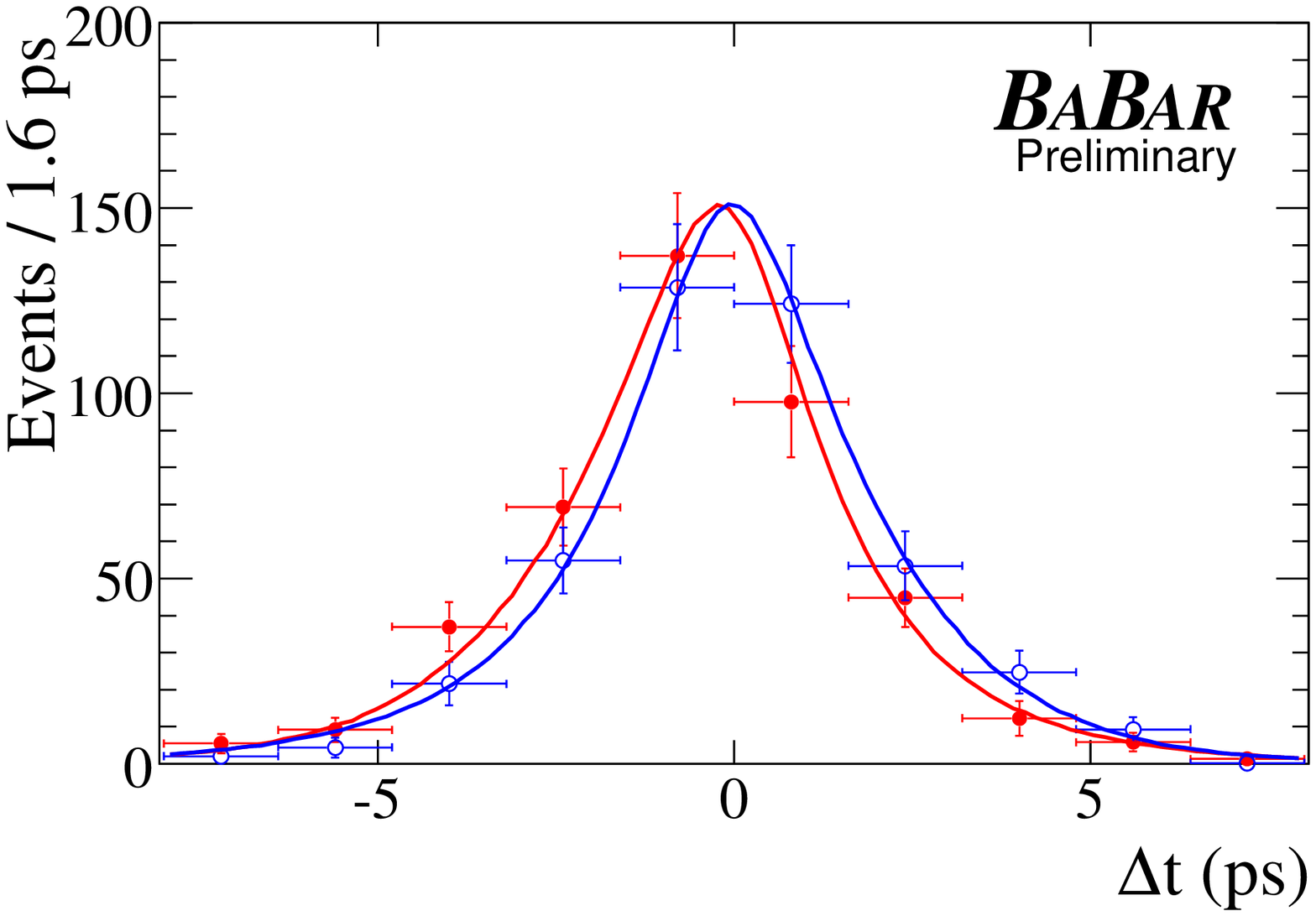}     &  \includegraphics[width=8cm,height=4cm]{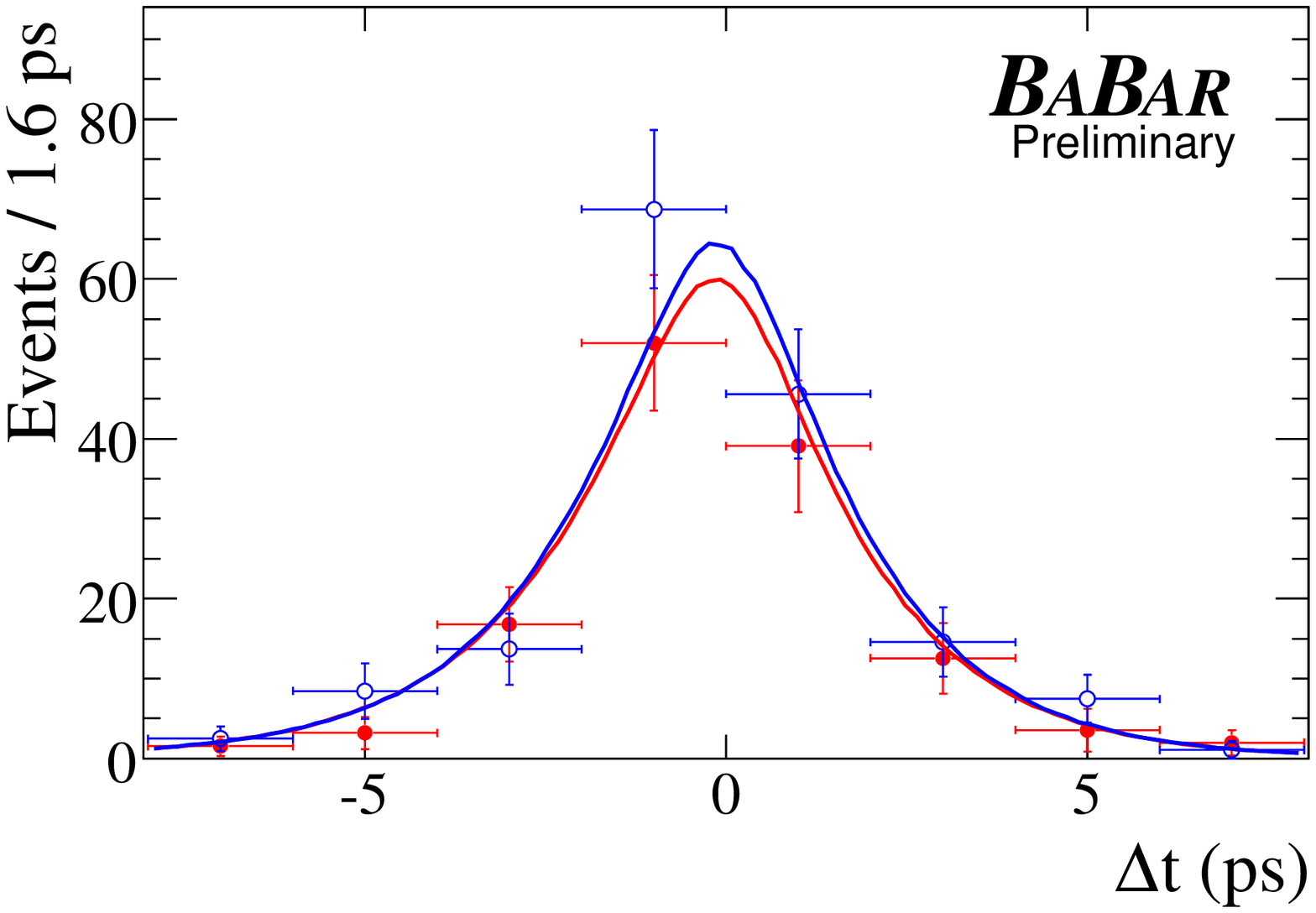}  \\
\includegraphics[width=8cm,height=4cm]{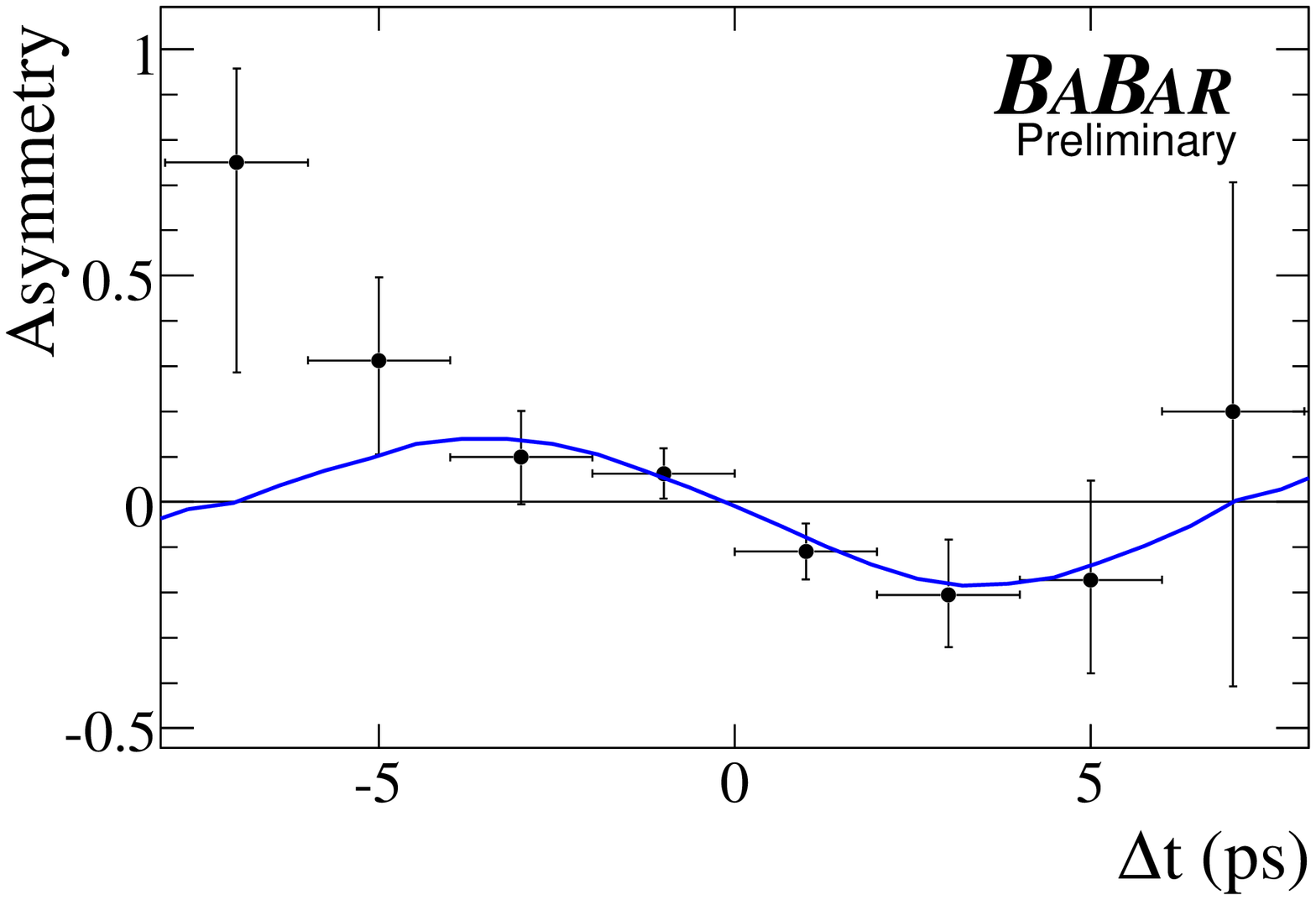} & \includegraphics[width=8cm,height=4cm]{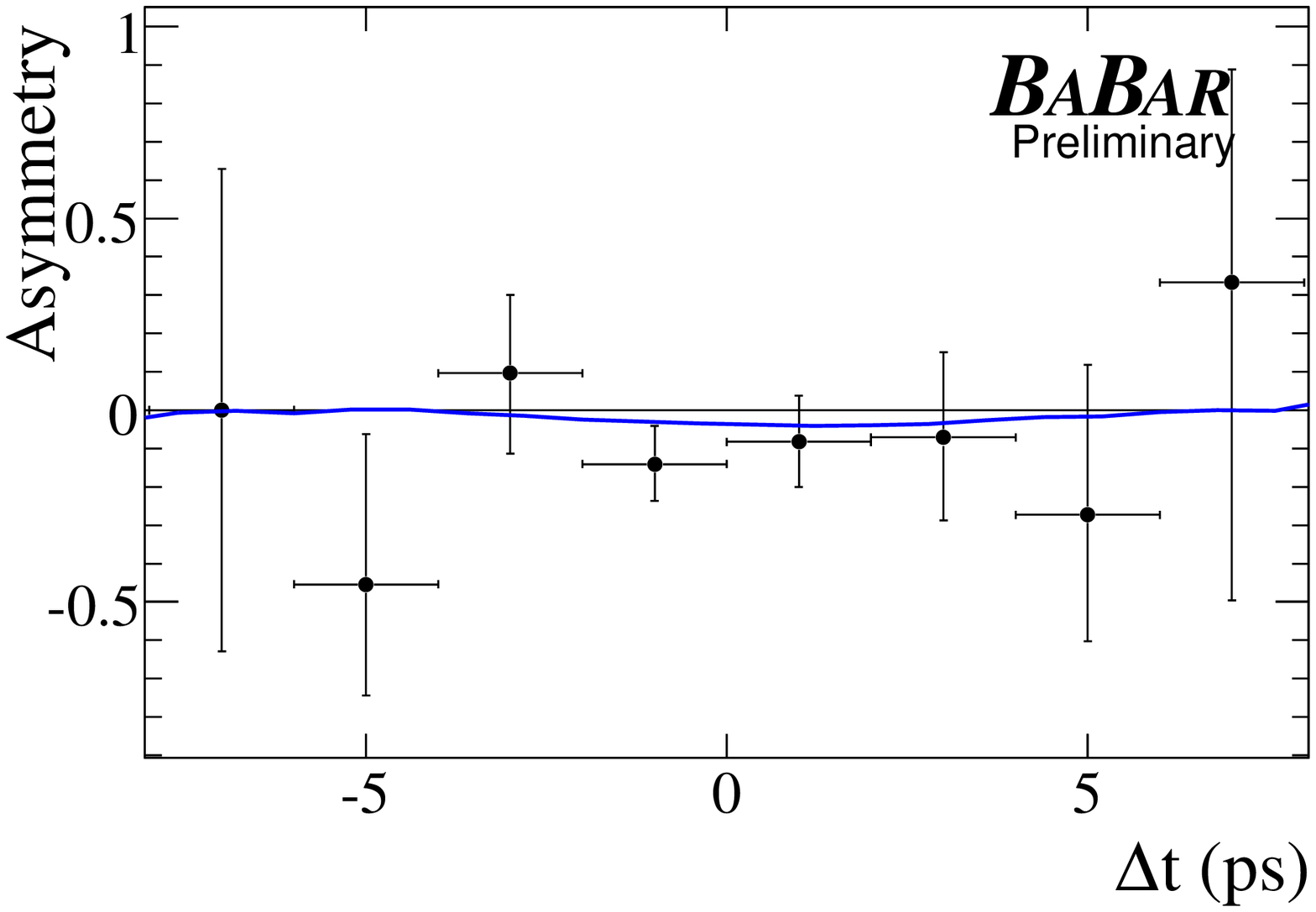}
\end{tabular}
\caption{ The \deltat\ (top)  distributions and asymmetries (bottom) in the whole DP (left) and \LowMass region (right),
for the \KKKspm mode.
For the \deltat\ distributions, \Bz- (\Bzb-) tagged signal-weighted events are shown as filled (open) 
circles, with the PDF projection in dashed red (solid blue).}
\label{fig:dt}
\end{figure}

We perform a fit to both 4316 $\Bz\to\KKKspm$ and 2205 $\Bz\to\KKKszz$ candidates simultaneously in the full Dalitz plot.
In this step we assume that all charmless decays have the same \CP-asymmetry parameters. A Fisher discriminant cut ($-2.5<\fisher<4$), which retains 
about 95\% of signal events and 60\% of continuum events, is applied. We do not include the Fisher PDF in the fit.
We vary the event yields, isobar coefficients, and the two
\CP-asymmetry parameters $\Acp$ and $\betaeff$ averaged over the Dalitz plot. 
We find a signal yield of $1268 \pm  43$ (\KKKspm) and $160 \pm 19$ (\KKKszz) events, and a \BB background yield of
$47 \pm 31$ (\KKKspm) and $24 \pm 16$ (\KKKszz) events.
The isobar amplitudes, phases, and fractions are listed in Table~\ref{tab:wholeDP}.  The resonant fractions do not add up 
to 100\% due to interference between the resonances. The \CP-asymmetry parameters, and the correlation coefficients 
$\rho$ between them, are summarized in Table~\ref{tab:acp}.  Fig.~\ref{fig:DV-wholeDP} shows a projection of the Dalitz 
plot variable \mKK. Fig.~\ref{fig:dt} shows distributions of $\Delta t$ for \Bz-tagged and \Bzb-tagged events, and the 
asymmetry ${\cal A}(\Delta t)=(N_{\Bz}-N_{\Bzb})/(N_{\Bz}+N_{\Bzb})$, obtained with the \splot technique.

\begin{figure}[ptb]
\center
\subfigure[]{\epsfig{file=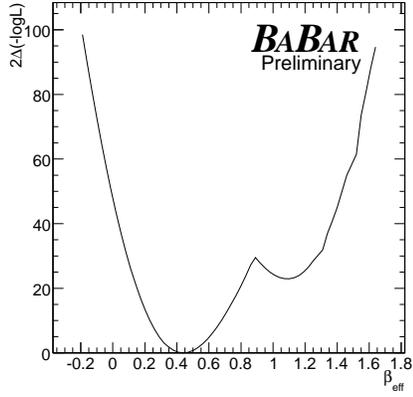, width=0.35\textwidth}} \hspace{0.1 in}
\subfigure[]{\epsfig{file=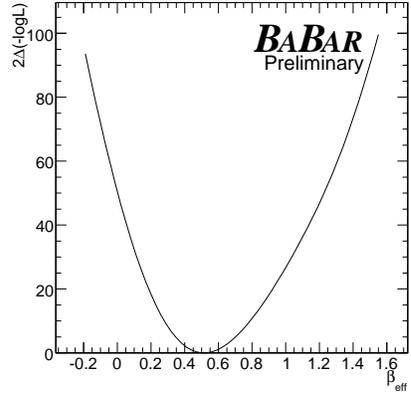, width=0.35\textwidth}}\\
\subfigure[]{\epsfig{file=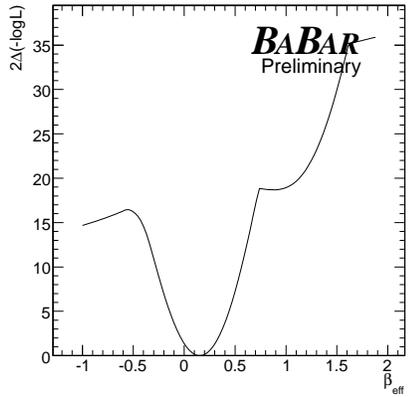, width=0.35\textwidth}} \hspace{0.1 in}
\subfigure[]{\epsfig{file=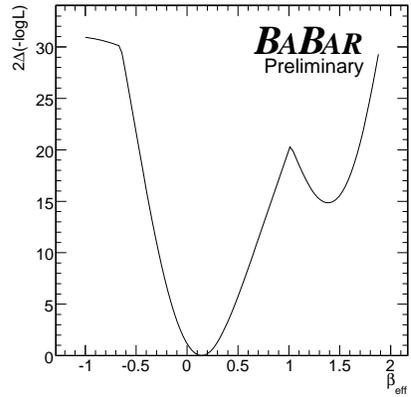, width=0.35\textwidth}}\\
\caption{ The change in the value of -2log($\mathcal L$) as a function of \betaeff, for (a) the whole DP, (b) the \HighMass region, (c) \fzone, and (d) \hjphi.}
\label{fig:scan}
\end{figure}

To calculate the significance of the nominal \betaeff result, many fits are performed with fixed but different \betaeff values. The change in  
likelihood as a function of \betaeff is shown in Fig.~\ref{fig:scan}.

\begin{table}[h]
\center
\begin{tabular}{|lr|rrr|}
\hline \hline
\multicolumn{2}{|l|}{Decay }            &     Amplitude $c_r$   &       Phase $\phi_r$       & Fraction ${\cal F}_r$ (\%) \\
\hline  
\multicolumn{2}{|l|}{$\phi(1020)\KS$}	& $  0.00897 \pm 0.00096$ &       $ -0.341 \pm  0.232$   & $12.6 \pm  1.0$            \\  
\multicolumn{2}{|l|}{$f_0(980)\KS$}	& $  0.542 \pm 0.044$  &       $ -0.201 \pm  0.157$   & $27.8 \pm  7.1$            \\ 
\multicolumn{2}{|l|}{$X_0(1550)\KS$}	& $  0.141 \pm 0.017$  &       $ -0.370 \pm  0.154$   & $5.70\pm  1.70$            \\ 
$NR$ &$(\Kp\Km)$      			& 1 (fixed)             &       0 (fixed)            & $98.1 \pm 18.7$            \\ 
        &$(\Kp\KS)$             	& $  0.328 \pm 0.058$     &       $  1.81  \pm 0.23$   & $10.5 \pm  3.4$            \\ 
        &$(\Km\KS)$             	& $  0.353 \pm 0.066$  	&       $ -1.44  \pm 0.27$   & $12.1 \pm  3.8$            \\  
\hline
\multicolumn{2}{|l|}{$\chi_{c0}\KS$}    & $  0.0298\pm 0.0046$  &       $ 0.732 \pm 0.437$     & $2.53 \pm 0.60$            \\ 
\multicolumn{2}{|l|}{$\Dp\Km$}          & $  1.34\pm 0.19$    	&               --           & $3.43 \pm 0.69$         	  \\ 
\multicolumn{2}{|l|}{$\Ds\Km$}          & $  0.826\pm 0.160$     	&               --           & $1.37 \pm 0.46$            \\ 
\hline \hline
\end{tabular}
\caption{Isobar amplitudes and phases from the fit to the whole DP. 
Three rows for non-resonant (NR) contribution correspond to coefficients of exponential functions in Eq.~(\ref{eq:nr}). The errors are statistical only. }
\label{tab:wholeDP}
\end{table}

\begin{table}[h]
\center
\begin{tabular}{|l|rr|}
\hline \hline
Name                    &      Whole DP         &      \HighMass \\
\hline
$\Acp$        & $ 0.03 \pm 0.07 \pm 0.02$    &  $0.05 \pm 0.09 \pm 0.04$  \\
\betaeff      & $ 0.44 \pm 0.07 \pm 0.02$    &  $0.52 \pm 0.08 \pm 0.03$ \\ 
$\rho$	      & $ 0.041$	                &       $0.031$ 	\\
\hline \hline
\end{tabular}
\caption{The \CP-asymmetry parameters from the whole DP fit and the \HighMass region fit. 
The first error is statistical and the second is systematic. $\rho$ is the correlation coefficient.}
\label{tab:acp}
\end{table}

\subsection{\HighMass fit}
\begin{figure}[ptb]
\center
\begin{tabular}{c}
\includegraphics[height=6.5cm]{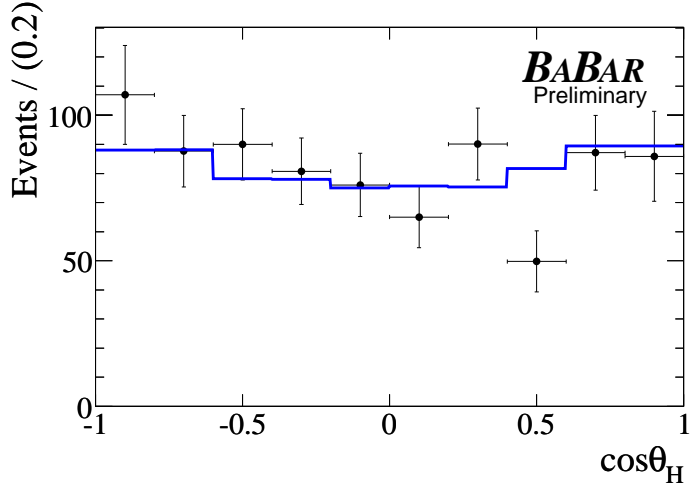} 
\vspace{-.5cm}
\end{tabular}
\caption{For the \HighMass region fit, the distribution of the Dalitz plot variable \cosH for 
signal-weighted data events (points) compared with the fit PDF in the \KKKspm mode.}
\label{fig:DV-HighMass}
\end{figure}

We perform a fit to both 3112 $\Bz\to\KKKspm$ and 1917 $\Bz\to\KKKszz$ candidates in the \HighMass region ($\mKK>1.1\gevcc$) simultaneously. We fix all isobar coefficients to the values from the whole DP fit. We vary yields and shared \CP-asymmetry parameters. We find a signal yield of $894 \pm 36$ \KKKspm and $117 \pm 16$ \KKKszz events, and a \BB background yield of
$50 \pm 31$ (\KKKspm) and $20 \pm 15$ (\KKKszz) events. The fit results are summarized in Table~\ref{tab:acp}. Fig.~\ref{fig:DV-HighMass} shows a projection of the Dalitz plot variable \cosH for events in this region, using the \splot technique.

\subsection{\LowMass fit}
In order to measure \CP-asymmetry parameters for components with low-\KpKm mass with reduced model-dependence from the rest of the DP, we select 
events using a cut of $\mKK<1.1\gevcc$. Because we are only selecting a small region of the DP, the correlation between the Fisher discriminant 
\fisher and the DP location is unimportant. We therefore relax the cut on \fisher, and add the \fisher PDF to the fit. After these requirements 
on \mKK and \fisher, there are 1846 (\KKKspm) and 493 (\KKKszz) candidates remaining. The most significant contributions in this region come 
from \hjphi\KS and \fzone\KS decays, with a smaller contribution from a low-\KpKm mass tail of non-resonant decays. We fix all the isobar 
coefficients except for those of the $\hjphi$ to the values from the whole DP fit, and fix the \CP-asymmetry parameters $b_r$ and $\delta_r$ 
for all resonances except the $\hjphi$ and $\fzone$ to be 0. 
We vary the events yields, isobar coefficients for the $\hjphi$, and separate \CP-asymmetry parameters for the $\hjphi$ and $\fzone$ in the fit. 
We find signal yields of $381 \pm 23$ (\KKKspm) and $40 \pm 9$ (\KKKszz) events, and \BB background yields of
$12 \pm 13$ (\KKKspm) and $-3 \pm 5$ (\KKKszz) events

 The \CP-asymmetry results are listed in Table~\ref{tab:low_mass_yields_cp}; the systematic uncertainties will be described in Sec.~\ref{sec:Systematics}. 
We find two solutions with likelihood difference $\Delta$log($\mathcal L$) = 0.1.  Solution (1) is consistent with the SM, while Solution (2)
has a value of \betaeff for the \fzone\KS decay that differs significantly from the SM, as shown in Table~\ref{tab:low_mass_yields_cp}.  The two
solutions also have significantly different values of $c_r$ for the $\hjphi$.  Both solutions also have a mathematical ambiguity of $\pm\pi$ radians
on \betaeff for the $\hjphi$, and a correlated ambiguity of $\pm\pi$ radians on the isobar parameter $\phi_r$ for the $\hjphi$.  This 
ambiguity is present because the decay amplitude contains interference terms that only depend on the linear combinations
$\betaeff+\phi_r$ and $\betaeff-\phi_r$.  We choose Solution (1) as our nominal solution.  The correlation coefficients $\rho$ between the 
\CP\ parameters for Solution (1) are shown in Table~\ref{tab:low_mass_yields_cp}.  Because the decay rate 
depends on interference terms between the \hjphi\KS and \fzone\KS decays, the significant correlation between the measured \CP parameters is 
expected.

\begin{table}[h]
\center
\begin{tabular}{|l|r|r|rrrr|}
\hline \hline
Name                            &     Solution (1)		&  Solution (2)  & \multicolumn{4}{c|}{Correlation}	\\

                                &                             	&                  &1	& 2	& 3 	& 4		\\
\hline
1 $\Acp(\phi\KS)$               &  $ 0.14 \pm 0.19 \pm 0.02 $   & $ 0.13 \pm 0.18$ &1.0	& -0.09 & -0.28  & 0.09		\\
2 $\betaeff (\phi\KS)$          &  $ 0.13 \pm 0.13 \pm 0.02 $   & $ 0.14 \pm 0.14$ &	& 1.0	& 0.54	& 0.65		\\
3 $\Acp(f_0\KS)$                &  $ 0.01 \pm 0.26 \pm 0.07 $  & $-0.49 \pm 0.25$ &	&	& 1.0	& 0.25 		\\
4 $\betaeff (f_0\KS)$           &  $ 0.15 \pm 0.13 \pm 0.03 $   & $ 3.44 \pm 0.19$ &    &	&	& 1.0		\\
\hline \hline
\end{tabular}
\caption{\CP-violation parameters for $\Bz\to\KKKs$ for $\mKK < 1.1~\gevcc$. The first error is statistical and the second is systematic.
   Correlation coefficients are given for Solution (1) only.}
\label{tab:low_mass_yields_cp}
\end{table}

\begin{figure}[ptb]
\center
\begin{tabular}{ll}
\includegraphics[height=5.5cm]{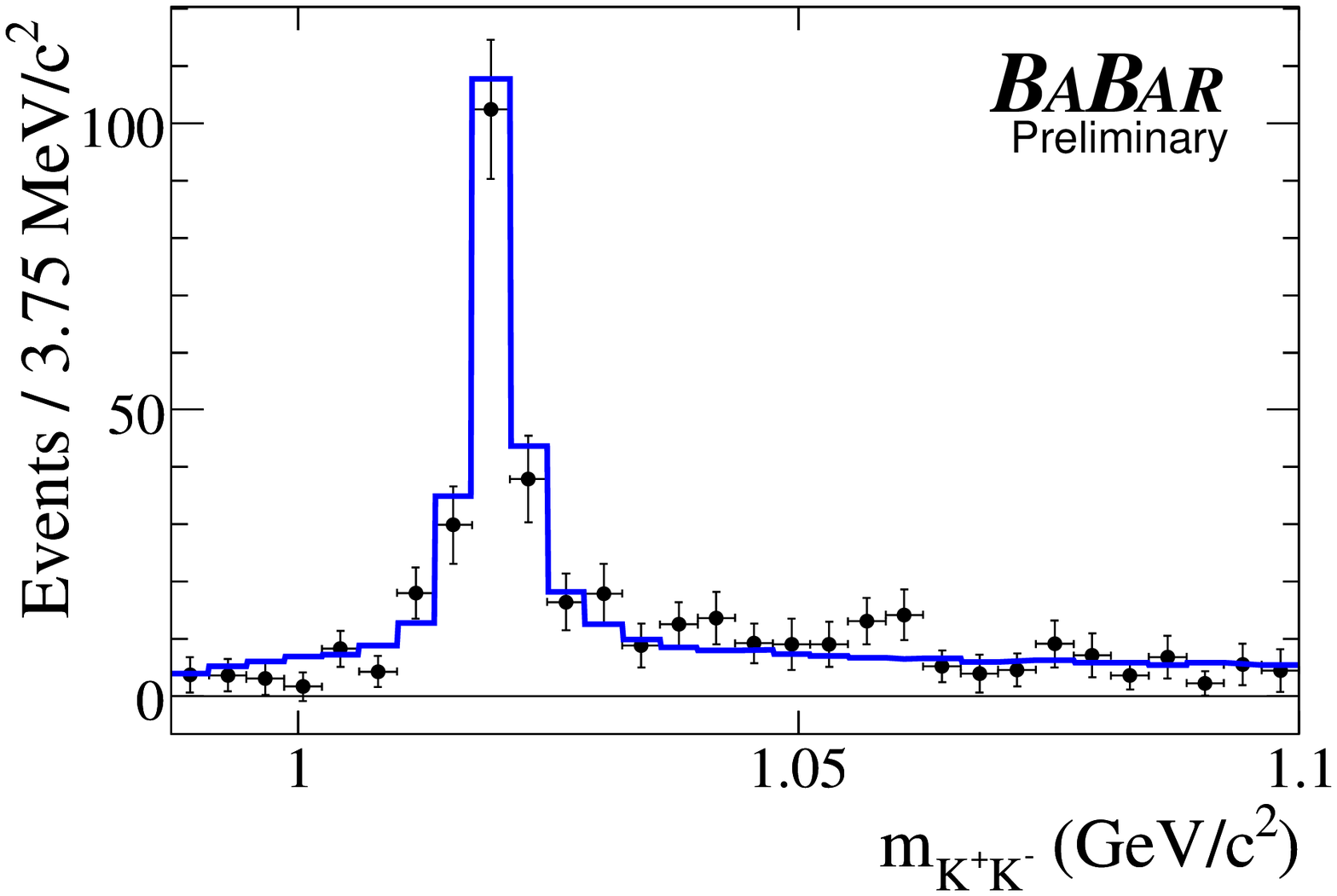} & \includegraphics[height=5.5cm]{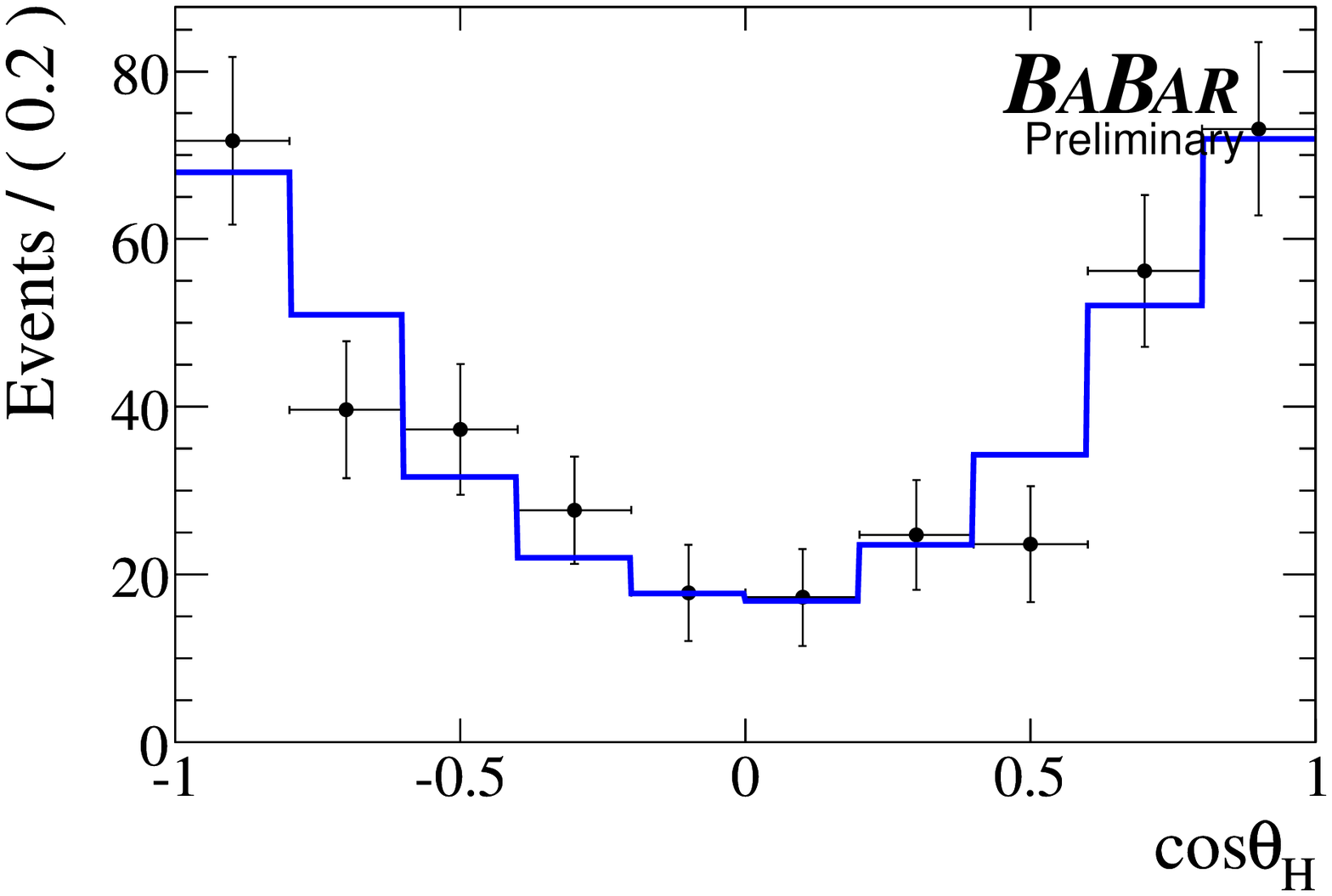} 
\end{tabular}
\caption{For the \LowMass region fit, the distributions of the Dalitz plot variables \mKK (left) and \cosH (right)
for signal-weighted data events (points) compared with the fit PDF in the \KKKspm mode.}
\label{fig:DV-LowMass}
\end{figure}

Fig.~\ref{fig:DV-LowMass} shows projections of the Dalitz plot distributions of events in this region, using the
\splot technique. Fig.~\ref{fig:dt} shows 
distributions of $\Delta t$ for \Bz-tagged and \Bzb-tagged events, and the asymmetry ${\cal A}(\Delta t)=(N_{\Bz}-N_{\Bzb})/(N_{\Bz}+N_{\Bzb})$. 

The decay $\Bz\to\hjphi\KS$, with highly suppressed tree amplitudes, is,
in terms of theoretical uncertainty, the cleanest channel to interpret
possible deviations of the \CP-violation parameters from the SM
expectations. 
Values of \betaeff are consistent with the value found 
in $[c\bar{c}]\Kz$ decays~\cite{Aubert:2004zt,Abe:2005bt}.

We also calculate the parameters $C$ and $-\eta S$ for $\hjphi\KS$ and $\fzone\KS$ using the expressions in (\ref{eq:C}) and (\ref{eq:etaS}). 
The results are shown in Table~\ref{tab:CandS}, along with $C$ and $-\eta S$ for the whole DP and \HighMass fits.

\begin{table}[h]
\center
\begin{tabular}{|l|c|c|}
\hline\hline
                  & $C$                     &     $-\eta S$       \\                   
\hline
Whole DP          &  $-0.03 \pm 0.07 \pm 0.02$       &  $0.77 \pm 0.09 \pm 0.02$  \\
\hline
\HighMass         &  $-0.05 \pm 0.09 \pm 0.04$       &  $0.86 \pm 0.08 \pm 0.03$   \\
\hline
$\hjphi\KS$       &  $-0.14 \pm 0.19 \pm 0.02$       & $0.26 \pm 0.26 \pm 0.03$    \\
$\fzone\KS$       &  $-0.01 \pm 0.26 \pm 0.07$       & $0.29 \pm 0.25 \pm 0.06$  \\
\hline\hline
\end{tabular}
\caption{The \CP asymmetry parameters $C$ and $-\eta S$, derived using Equations (\ref{eq:C}) 
and (\ref{eq:etaS}). Results are shown for the whole DP, the \HighMass region, and for both 
$\hjphi\KS$ and $\fzone\KS$ in the \LowMass region. For the \LowMass results, only Solution
(1) is shown. The first error is statistical and the second is systematic.\label{tab:CandS}}
\end{table}

\section{SYSTEMATIC STUDIES}
\label{sec:Systematics}

We study systematic effects on the \CP-asymmetry parameters
due to fixed parameters in the \mes and \DeltaE PDFs.
We assign systematic errors  by comparing the fit with nominal parameters and 
with parameters varied by their error ($\pm 1 \sigma$), and assign the average difference as the systematic error.
In addition, we account for a potential fit bias using values observed in studies 
with MC samples generated with the nominal  Dalitz plot model.
We take the average values of the bias observed in these studies as the systematic error.
We account for fixed \deltat\ resolution parameters, \Bz\ lifetime,
\Bz-\Bzb mixing and flavor tagging parameters. We also assign an error
due to interference between the CKM-suppressed $\bar{b}\to\bar{u} c\bar{d}$
and the favored $b\to c\bar{u}d$ amplitude for some tag-side $B$ decays~\cite{dcsd}.
Smaller errors due to beam-spot position uncertainty, detector alignment, and the
boost correction are based on studies done in charmonium decays.  In the cases of the \LowMass
and \HighMass fits, we also assign systematic errors due to the isobar coefficients that are
fixed to the result from the whole DP fit.
In all fits we assume no direct \CP\ violation in decays dominated by the $b\to c$ transition 
($\chi_{c0}\KS$, $D_{(s)}K$).

We also assign an error due to uncertainty in the resonant and non-resonant
line-shape parameters. The systematic uncertainty associated with the resonant component includes
the uncertainty in the mass and width of the X(1550), estimated by replacing the parameters used in
the nominal fit with the values found by different measurements: $m_r=1.491$~\gevcc, 
$\Gamma=0.145$~\gev~\cite{Garmash:2004wa}. 
All the systematic uncertainties are summarized in Table.~\ref{tab:sys_errors}.

\begin{table}[h]
\center
\begin{tabular}{|l|rr|rrrr|rr|}
\hline \hline
Parameter &          \multicolumn{2}{c|}{Whole DP}  &    \multicolumn{2}{c}{$\phi\KS$} &         \multicolumn{2}{c|}{$f_0\KS$}   &       \multicolumn{2}{c|}{\HighMass}    \\
                        & $\Acp$       &  \betaeff        &  $\Acp$    &  \betaeff &   $\Acp$   &  \betaeff  &   $\Acp$   &  \betaeff        \\ \hline 
\hline
Fixed PDF Parameters    &  0.010 & 0.010   &  0.014 & 0.010   & 0.025 & 0.015   & 0.013 & 0.010           \\
Fit Bias                &  0.007 & 0.011   &  0.009 & 0.012   & 0.011 & 0.011   & 0.014 & 0.009           \\
DCSD, Beam Spot, other  &  0.015 & 0.004   &  0.015 & 0.004   & 0.015 & 0.004   & 0.015 & 0.004           \\
Dalitz Model            &  0.005 & 0.005   &  0.009 & 0.002   & 0.060 & 0.024   & 0.027 & 0.023           \\ \hline
Total                   &  0.020 & 0.016   &  0.024 & 0.016   & 0.068 & 0.031   & 0.036 & 0.026           \\
\hline \hline
\end{tabular}
\caption{Summary of systematic errors on \CP-asymmetry parameters.  Errors for 
$\phi\KS$ and $f_0\KS$ \CP-parameters are based on the \LowMass region fit. Total is obtained from the quadratic sum of the individual systematics.}
\label{tab:sys_errors}
\end{table}

\section{CONCLUSIONS}
\label{sec:Summary}

We performed a ML fit to analyze the DP distribution of $\Bz \to \KKKs$ decay  with the full \babar\ dataset. From a
fit to the whole DP, we measure $\betaeff=0.44\pm 0.07 \pm 0.02$, $\Acp=0.03\pm 0.07\pm 0.02$, consistent
with our previous measurements~\cite{Previous} and compatible with the Standard Model values $\beta \simeq 0.37, \Acp = 0$.
We measure \CP violation with a significance of 6.7 standard deviations (including statistical and systematic errors), 
and we reject the solution near $\pi/2 - \beta$ at 4.8 standard deviations.   

From a fit to the region of the DP with $\mKK>1.1\gevcc$, we measure $\betaeff= 0.52 \pm 0.08 \pm 0.03$ and 
$\Acp= 0.05 \pm 0.09  \pm 0.04$, compatible with the Standard Model expectations.  We measure \CP violation
in this \HighMass region at 6.7 standard deviations.

From a fit to events at low \Kp\Km masses, we measure $\betaeff= 0.13 \pm 0.13 \pm 0.02$ and $\Acp= 0.14 \pm 0.19  \pm 0.02$
for $\Bz \to \phi(1020)\KS$, and $\betaeff= 0.15 \pm 0.13 \pm 0.03$ and $\Acp= 0.01 \pm 0.26 \pm 0.07$ for 
$\Bz \to f_0\KS$.  The results for $\betaeff$ are roughly 1.7 standard deviations below the Standard Model value.

These results supersede our previous measurements~\cite{Previous} made on a smaller dataset.  All of our
results are consistent with our previous measurements.

\section{ACKNOWLEDGMENTS}
\label{sec:Acknowledgments}

We are grateful for the 
extraordinary contributions of our \pep2\ colleagues in
achieving the excellent luminosity and machine conditions
that have made this work possible.
The success of this project also relies critically on the 
expertise and dedication of the computing organizations that 
support \babar.
The collaborating institutions wish to thank 
SLAC for its support and the kind hospitality extended to them. 
This work is supported by the
US Department of Energy
and National Science Foundation, the
Natural Sciences and Engineering Research Council (Canada),
the Commissariat \`a l'Energie Atomique and
Institut National de Physique Nucl\'eaire et de Physique des Particules
(France), the
Bundesministerium f\"ur Bildung und Forschung and
Deutsche Forschungsgemeinschaft
(Germany), the
Istituto Nazionale di Fisica Nucleare (Italy),
the Foundation for Fundamental Research on Matter (The Netherlands),
the Research Council of Norway, the
Ministry of Education and Science of the Russian Federation, 
Ministerio de Educaci\'on y Ciencia (Spain), and the
Science and Technology Facilities Council (United Kingdom).
Individuals have received support from 
the Marie-Curie IEF program (European Union) and
the A. P. Sloan Foundation.

\end{document}